# Spreadsheet End-User Behaviour Analysis

Brian Bishop, Kevin McDaid
Dundalk Institute of Technology,
Dundalk, Ireland
brian.bishop@dkit.ie, kevin.mcdaid@dkit.ie

## ABSTRACT

*To aid the development of spreadsheet debugging tools, a knowledge of end-users natural behaviour within the Excel environment would be advantageous. This paper details the design and application of a novel data acquisition tool, which can be used for the unobtrusive recording of end-users mouse, keyboard and Excel specific actions during the debugging of Excel spreadsheets. A debugging experiment was conducted using this data acquisition tool, and based on analysis of end-users performance and behaviour data, the authors developed a 'spreadsheet cell coverage feedback' debugging tool. Results from the debugging experiment are presented in terms of end-user debugging performance and behaviour, and the outcomes of an evaluation experiment with the debugging tool are detailed.*

## 1. INTRODUCTION

The reported usage of spreadsheet programs spans a wide variety of job functions, purposes and industries. In a survey of nearly 1600 respondents, Baker et al [2006] found that spreadsheets were used by end-users in various job functions including finance, engineering, manufacturing, marketing, sales and administration, and for many different purposes, such as maintaining lists, analysing and tracking data and determining trends. Iyengar & Svirbely [2005] reported on the usage of a website that made available medical algorithms in the form of MS Excel files, and the users of the website included physicians, nurses, healthcare professionals, computer scientists etc. Maybe more than any other industry, spreadsheets are of critical importance to the finance sector [Croll, 2005]. In a study on the use of spreadsheets in organisations in the City of London, Croll [2005] found that with regard to the financial markets:

*"Excel is utterly pervasive. Nothing large (good or bad) happens without it passing at some time though Excel."*

The most recent study that investigated spreadsheet error rates was conducted by Powell, Baker & Lawson [2007], who reported that of the 50 real-world operational spreadsheets they audited, 94% contained errors. To help alleviate the problem of spreadsheet errors, a number of tools are available to aid in the debugging of spreadsheet programs. These tools are both academic [Burnett et al, 2002], [Abraham & Erwig, 2007], [Clermont & Mittermeir, 2003] and commercial products such as Spreadsheet Professional (www.spreadsheetinnovations.com) and RedRover Audit (www.redroversoftware.com). In order to develop tools that complement end-users natural behaviour, some knowledge of that behaviour would be required. The concept of human-centered development in software engineering could be applied to the development of spreadsheet debugging tools. Norman [1999] stated that *"At its core, human-centered product development requires developers who understand people and the tasks they wish to achieve. It means starting by observing and working with users".* At present, there is very little empirical data available to researchers on the processes and actions of end-users while debugging Excel spreadsheets.

This paper details the design and application of a novel custom built data acquisition tool. An experiment was conducted with 47 subjects (professionals and students), the aim of which was to record and analyse the performance and behaviour of expert and novice end-users while debugging an experimental spreadsheet model. Using the data gathered by the data acquisition tool, analysis was conducted which led the authors to develop a simple 'spreadsheet cell coverage feedback' debugging tool. Results from the experiment are presented in terms of end-user debugging performance and behaviour, and the outcomes of an evaluation experiment with the debugging tool are detailed.

The layout of the paper is as follows. Section 2 details data acquisition methods currently available and commonly used, and the custom built data acquisition tool developed by the authors. In Section 3, an experiment in which the data acquisition tool was utilised is described, and results are presented in terms of expert and novice performance and behaviour. The debugging tool developed by the authors is detailed in Section 4, along with results of an evaluation of the debugging tool which involved further use of the data acquisition tool. A conclusion to the paper is in Section 5.

## 2. DATA ACQUISITION IN SPREADSHEET RESEARCH

A number of methods are available for acquiring human computer interaction (HCI) data during experiments. These include video recording, screen recording, thinking-aloud protocol and eye-gaze tracking. These methods require subjects to be onsite with the researcher, or at the very least for the subject to install some recording software or equipment on their pc or within their working environment. A problem with data capture when using the somewhat intrusive data acquisition methods mentioned, is that a person's behaviour changes when they are aware that they are being observed. This phenomenon is commonly referred to as the observer effect, and also as the Panopticon effect or the Heisenberg effect [Liffick & Yohe, 2001].

To lessen the effects associated with using intrusive data acquisition methods, a mouse-and-keystroke type recording tool was developed by the authors, which could be embedded within an Excel spreadsheet, and as such be used remotely by subjects without the need for the author's presence. One of the main advantages of recording mouse-and-keystroke induced UI (user interaction) events is that they "*provide excellent data for quantitatively characterising on-line (on-screen) behaviour*" [Hilbert & Redmiles, 2000], and with this type of non-intrusive system monitoring the "*influence on participants by observation is zero*" [Spannagel, Gläser-Zikuda & Schroeder, 2005]. The authors believe that the influence on participants would not be zero, as subjects should be clearly informed of the data capture method being used, but nevertheless, the influence would be considerably lower than that of video recording and think-aloud methods. The next section details the design and workings of the data acquisition tool developed by the authors.

### 2.1 Custom Built Data Acquisition Tool: T-CAT

The authors developed a 'time-stamped cell activity tracking tool' (T-CAT) in VBA, which makes use of MS Excel's macro programming environment. The main advantages of developing T-CAT in VBA are: 1) The tool could be embedded in experimental spreadsheet models with no software installation required by participants and 2) The tool could easily access MS Excel's event listeners such as the Workbook functions: Open, BeforeClose, SheetActivate, SheetChange, and SheetSelectionChange.

A flowchart for T-CAT can be seen in Figure 1. The tool was designed to record the time and detail of all cell selection and cell change actions of individuals when debugging a spreadsheet. The T-CAT tool 'listens' for cell activity events, such as: worksheet selections, cell selections and cell edits. When an event occurs, the tool records all the details associated with the event, and stores them to arrays. When the spreadsheet is closed, MS Excel's (VBA's) BeforeClose event is initiated, and the values stored in the details arrays are printed to a hidden worksheet. A popup message box then requests that the user email the spreadsheet to the researcher.

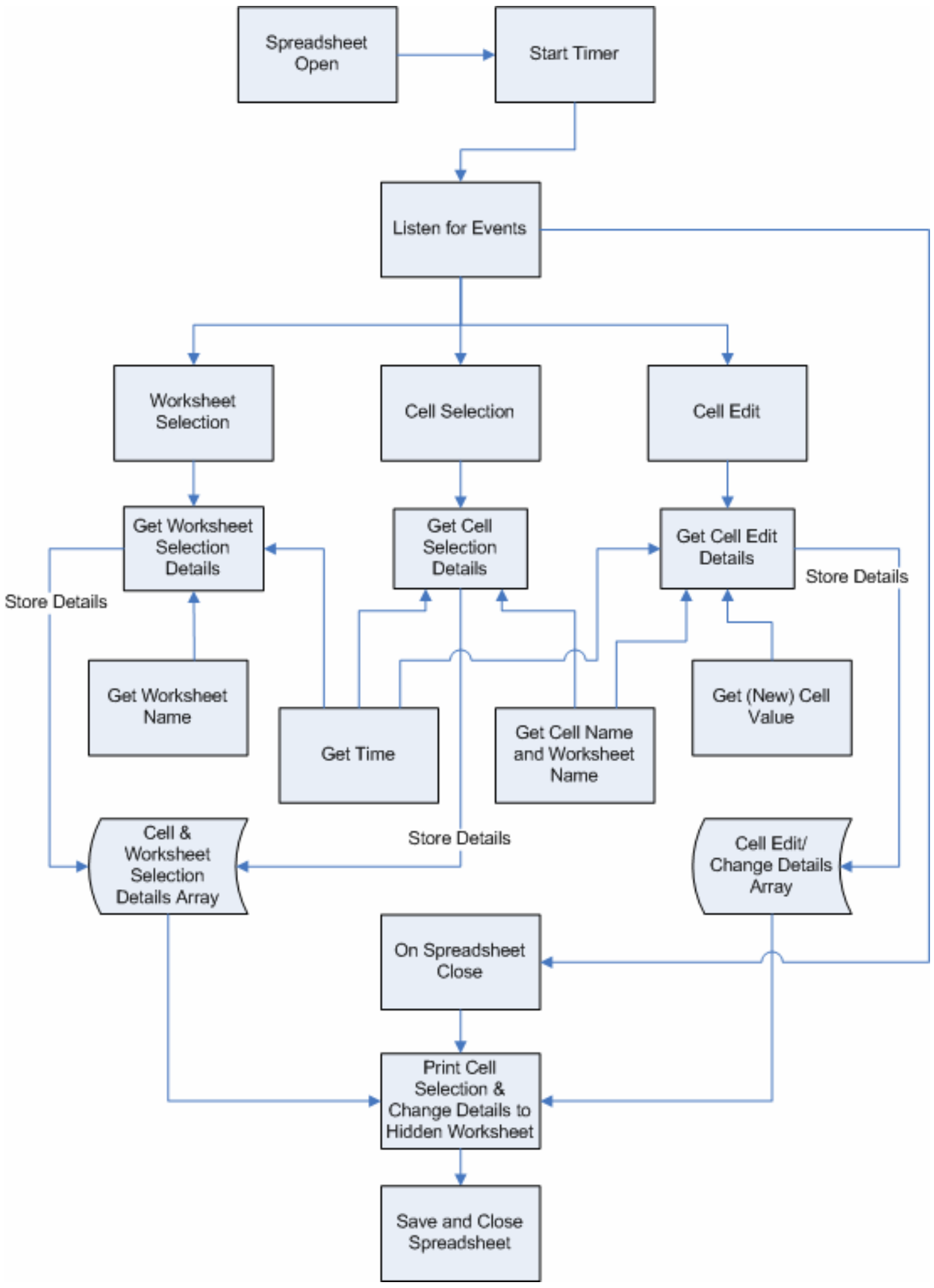

**Figure 1: Flowchart for T-CAT**

The data recorded by T-CAT during the debugging process is as follows: Individual cells selected, cell ranges selected, worksheet selections, individual cells edited and the resulting cell or formula values, cell ranges changed and resulting cell or formula values. Timestamps are recorded for all of the above (in seconds and milliseconds). More complex spreadsheet activities can also be identified from the resulting data log, including copy and paste, undo typing, redo typing and drag-and-fill. It should be noted that the T-CAT tool does not record the use of native Excel debugging aids such as switching to Formula View or Tracing Precedents/Dependents. As mentioned earlier, the data recorded by T-CAT is printed to a hidden worksheet within the experimental spreadsheet when the spreadsheet is closed. Figure 2 shows a sample of debugging data recorded by T-CAT.

Looking at cell ‘A3’ from Figure 2, the value shows that the cell ‘D10’ on the ‘Payroll’ worksheet was edited after 40.828125 seconds, with a resulting cell value of ‘40’. The process of editing this cell can be determined by looking at cells ‘E/F32’ and ‘E/F33’ in the same figure; ‘Payroll D10’ was selected after 33.046875 seconds, was edited, and then the Return key was pressed bringing the control to ‘Payroll D11’. The move of cell focus from ‘Payroll D10’ to ‘Payroll D11’ (as can be seen from cells ‘E32 - E33’ in the figure) shows that the cell edit was completed after 40.828125 seconds elapsed time. This gives a time for editing cell ‘Payroll D10’ of 7.78125 seconds.

| | A | B | C | D | E | F | G |
|---|---|---|---|---|---|---|---|
| 1 | Cell Edit and Resulting Cell Value | Time (Seconds) | | | Cell Selcecion Details | Time (Seconds) | |
| 2 | | | | | | | |
| 3 | Payroll$D$10 40 | 40.828125 | | | Payroll | 0 | |
| 4 | Payroll$D$11 40 | 43.15625 | | | Payroll$C$5 | 22.953125 | |
| 5 | Payroll$E$14 10 | 72.359375 | | | Payroll$C$6 | 23.75 | |
| 6 | Payroll$F$9 =C9*D9 | 87.359375 | | | Payroll$C$7 | 25.546875 | |
| 7 | Payroll$F$10 =C10*D10 | 96.5 | | | Payroll$C$8 | 25.84375 | |
| 8 | Payroll$C$13 6.75 | 134.890625 | | | Payroll$C$9 | 26.09375 | |
| 9 | Payroll$F$12:$F$14 =C12*D13 | 148.796875 | | | Payroll$C$10 | 26.3125 | |
| 10 | Payroll$F$10:$F$14 =C10*D10 | 157.375 | | | Payroll$C$11 | 26.484375 | |
| 11 | Payroll$G$5 0 | 196.34375 | | | Payroll$C$12 | 26.65625 | |
| 12 | Payroll$G$6 0 | 196.859375 | | | Payroll$C$13 | 26.828125 | |
| 13 | Payroll$G$7 0 | 197.5 | | | Payroll$C$14 | 27.03125 | |
| 14 | Payroll$G$11:$G$14 =C11*1.5*E11 | 229.375 | | | Payroll$C$13 | 28.25 | |
| 15 | Payroll$G$16 =SUM(G5:G14) | 238.046875 | | | Payroll$C$12 | 28.4375 | |
| 16 | Payroll$H$5 =F5+G5 | 263.3671875 | | | Payroll$C$11 | 28.609375 | |
| 17 | Payroll$H$16 =SUM(H5:H14) | 278.0703125 | | | Payroll$C$10 | 28.765625 | |
| 18 | Payroll$I$10 =SUM(H8:H10) | 300.9140625 | | | Payroll$C$9 | 28.9375 | |
| 19 | Payroll$I$14 =SUM(H11:H14) | 306.6484375 | | | Payroll$C$8 | 29.09375 | |
| 20 | Payroll$J$14 =SUM(I7:I14) | 320.4921875 | | | Payroll$C$7 | 29.265625 | |
| 21 | Payroll$J$16 =SUM(H5:H14) | 334.5859375 | | | Payroll$C$6 | 29.453125 | |
| 22 | Payroll$J$16 | 336.9296875 | | | Payroll$C$5 | 29.609375 | |
| 23 | Payroll$J$14 | 337.5546875 | | | Payroll$D$5 | 30.109375 | |
| 24 | Payroll$E$5 0 | 383.9296875 | | | Payroll$D$6 | 30.921875 | |
| 25 | Payroll$C$16 =SUM(C5:C14) | 428.0234375 | | | Payroll$D$7 | 31.109375 | |
| 26 | Payroll$C$16 =SUM(C5:C14) | 428.9765625 | | | Payroll$D$8 | 31.328125 | |
| 27 | Payroll$C$16 =SUM(C5:C14) | 435.4609375 | | | Payroll$D$9 | 31.515625 | |
| 28 | Payroll$C$16 | 436.1328125 | | | Payroll$D$10 | 31.703125 | |
| 29 | Office Expenses$D$12 Mar | 453.7578125 | | | Payroll$D$11 | 31.890625 | |
| 30 | Office Expenses$D$16 350 | 461.3984375 | | | Payroll$D$12 | 32.09375 | |
| 31 | Office Expenses$D$17 450 | 463.1328125 | | | Payroll$D$11 | 32.78125 | |
| 32 | Office Expenses$F$18 =SUM(F13:F17) | 498.6953125 | | | Payroll$D$10 | 33.046875 | |
| 33 | Office Expenses$C$5 261 | 555.9140625 | | | Payroll$D$11 | 40.828125 | |
| 34 | Office Expenses$F$5 =4*E5 | 567.1015625 | | | Payroll$D$12 | 43.15625 | |
| 35 | Office Expenses$E$8 =SUM(B8:D8) | 595.6171875 | | | Payroll$D$13 | 44.171875 | |

**Figure 2: Sample of T-CAT Recorded Spreadsheet Debugging Data**

## 3. T-CAT APPLIED IN A SPREADSHEET DEBUGGING EXPERIMENT

The T-CAT data acquisition tool was an integral part of a study [Bishop & McDaid, 2007] undertaken by the authors that aimed to record and analyse the performance and behaviour of 47 professional and student spreadsheet users while debugging a spreadsheet model seeded with errors. Participants were instructed to debug the spreadsheet, and each error found was to be corrected directly in the spreadsheet itself. The spreadsheet model was adapted from a model used in [Howe & Simkin, 2006]. The names and functions of the spreadsheet’s three worksheets were as follows: *Payroll*, compute typical payroll expenses; *Office Expenses*, compute office expenses; *Projections*, perform a 5-year projection of future expenses. Each worksheet had different error characteristics. *Payroll* had data entry, rule violation and formula errors; *Office Expenses* had clerical, data entry and formula errors; *Projections* had mostly formula errors. The spreadsheet model, experiment methodology and sample are presented and described in detail in [Bishop & McDaid, 2007]. The updated findings from this experiment, which relied heavily on the data captured by T-CAT, are presented below in terms of debugging performance, debugging behaviour, and overall cell coverage.

### 3.1 Overall Debugging Performance

Formal tests of significance for the difference between the results of student and professional debugging performance were conducted. One sided hypothesis tests, based on the Student T distribution and examining whether the mean performance of professionals exceeded the mean performance of students in discovering the four types of errors, were conducted using a significance level of 5%. Given the relatively low number of professional subjects and the uncertainty as to the nature of the distribution of the number of errors found by individual participants the one sided tests of a difference in the measure of centrality were also conducted using the non parametric Mann-Whitney U-test.

| Error Type | Number of Seeded Errors | % Corrected: Professionals | % Corrected : Students | T-test P-value | Mann Whitney P-value |
|---|---|---|---|---|---|
| Clerical/Non-Material | 4 | 17% | 11% | 0.15106 | 0.1532 |
| Rules Violation | 4 | 85% | 65% | 0.02421* | 0.0047** |
| Data Entry | 8 | 68% | 63% | 0.10121 | 0.17376 |
| Formula | 26 | 79% | 63% | 0.00031*** | 0.0086** |
| **Total** | 42 | 72% | 58% | 0.000063*** | 0.00131** |

**Figure 3: Debugging Performance – Professional and Student**

From the evidence of the overall results for students and professionals given in Figure 3 and the distinctions between the two groups for each error type, the professional group has been shown to outperform the student group. To determine if the overall difference between the two groups was statistically significant, t-tests were performed on the number of errors corrected for both the professionals and students, see Figure 3. The P-value of 0.000063, along with a Mann-Whitney U-test P-value of 0.001313, indicates that professionals outperform students in spreadsheet debugging and that the difference is statistically significant. The expert spreadsheet users also completed the debugging of the spreadsheet in a shorter time. Expert subjects completed the task in an average of 28 minutes, while the novice subjects completed the debugging task in an average 36 minutes. Overall, the expert spreadsheet users were found to be more efficient and effective spreadsheet debuggers than the novice spreadsheet users. They were found to be significantly better at debugging Rule Violation and Formula errors.

### 3.2 Debugging Behaviour

In order to determine the main areas within the spreadsheet that the participants focused on during the debugging experiment, a '*coverage per cell'* data analysis tool was developed which calculates, for each cell in the spreadsheet, the percentage of participants that inspected that cell. There were some difficulties in quantitatively representing professional and student participant's main areas of spreadsheet cell focus in a manner that would be intuitive and meaningful. Given that the human visual system is very good at clustering and recognising patterns and trends in data visualisations [Bealle, 2007], the data generated during the *coverage per cell* analysis was collated and represented in the form of colour coding within each of the three worksheets that made up the experimental spreadsheet. The colour coding made it easier to recognise the groups of cells that participants inspected and the difference between the student and professional participant's behaviour. An example of the colour coding can be seen for the Projections worksheet in Figure 5 and Figure 6 (Professional and Student coverage respectively), in which the individual cells are colour coded to show what percentage of the participants inspected them. Cells that contained column or row headings have been left unmarked, so the structure of the spreadsheet is clear. Cells that contained formulas, data or seeded errors are represented as follows:

- F – cell contains a formula.
- D – cell contains a data value (numeric data value).
- C Error – cell contains a seeded *clerical* error.
- RV Error – cell contains a seeded *rule violation* error.
- D Error – cell contains a seeded *data entry* error.
- F Error – cell contains a seeded *formula* error.

This analysis was carried out separately for both the student and professional subjects with the minimum time specified for a cell to be considered inspected set at 0.3 seconds. The minimum specified time of 0.3 seconds was set based on analysis of the minimum time it took participants to inspect a cell. The colour coding key can be seen in Figure 4. The bands of colour represent the percentage of participants that inspected each cell.

| | | |
|---|---|---|
| 90% - 100% | | Brown |
| 80% - 90% | | Orange |
| 70% - 80% | | Light Orange |
| 60% - 70% | | Light Yellow |
| 50% - 60% | | Rose |
| 40% - 50% | | Light Blue |
| 30% - 40% | | Sky Blue |
| 20% - 30% | | Pale Blue |
| 10% - 20% | | Light Turquoise |
| >0% - 10% | | Light Gray |

**Figure 4: Coverage per Cell Colour Coding Key**

It was found that for all worksheets and both participant groups, formula cells took precedence over data cells, but a greater percentage of professional participants looked at each cell. Summation formula cells and bottom-line value cells received more attention than other formula cells. Formula cells that outputted text values were not inspected as much as formula cells that outputted numeric values. This was the case for both the professionals and students.

A key finding concerned the inspection rates for some of the logically equivalent groups of formula cells. Participants checked the first cells in these groups, either the topmost cells in the case of vertically oriented groups or the leftmost cells for horizontally oriented groups, and a distinct *drop-off in inspection rates* could be seen for the rest of the cells in these logically equivalent groups. This was the case for both the professional and student participants, but this behaviour was more evident for the student (novice) sample.

| | A | B | C | D | E | F | G |
|---|---|---|---|---|---|---|---|
| 1 | C Error | | | | | | |
| 2 | Today's Date | C Error | | | | | |
| 3 | | | | | | | |
| 4 | | | | | | | |
| 5 | Projection Rates: | | | | | | |
| 6 | Payroll A Employees | D | | | | | |
| 7 | Payroll B Employees | D | | | | | |
| 8 | Payroll C Employees | D | | | | | |
| 9 | Office Variable Expenses | D | | | | | |
| 10 | Office Fixed Expenses | D | | | | | |
| 11 | | | | | | | |
| 12 | | | | | | | |
| 13 | | Current Year | Year 1 | Year 2 | Year 3 | Year 4 | Year 5 |
| 14 | **Payroll Expenses** | | | | | | |
| 15 | Category A | F | F | F | F | F | F |
| 16 | Category B | F | F | F | F | F | F |
| 17 | Category C | F Error | F | F | F | F | F Error |
| 18 | **Office Expenses** | | | | | | |
| 19 | Fixed Expenses | F Error | F Error | F Error | F Error | F Error | F Error |
| 20 | Variable Expenses | F Error | F | F | F | F | F |
| 21 | | | | | | | D Error |
| 22 | Total: | F | F | F | F | F | F Error |
| 23 | | | | | | | |
| 24 | Does yearly increase in expenses exceed 4% | | F | F | F | F | F |

**Figure 5: Professionals - Projections Sheet Coverage**

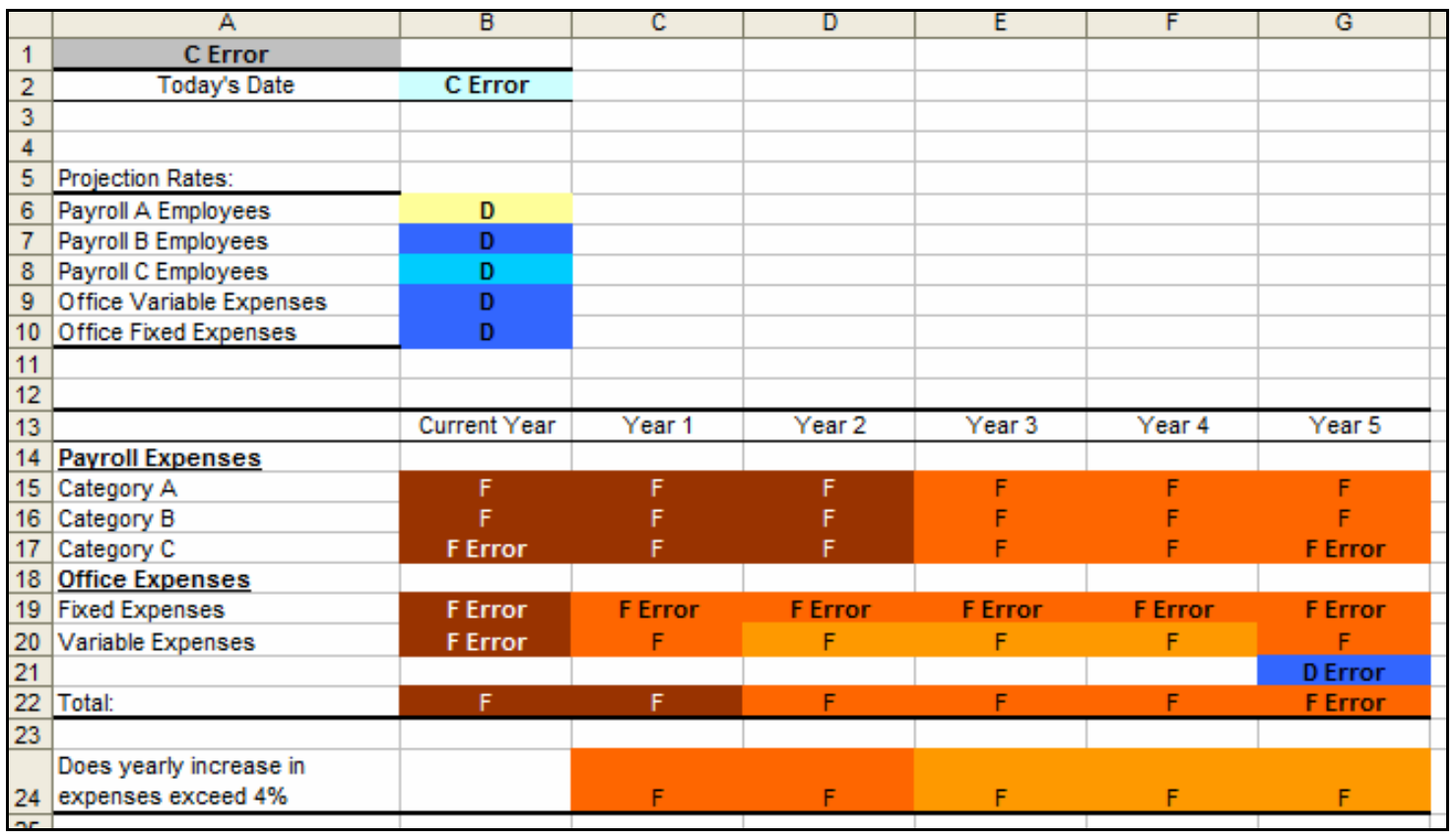

**Figure 6: Students - Projections Sheet Coverage**

## 3.3 Overall Cell Coverage

In software testing the quality of test suites is often measured through coverage-based measures which are considered to relate closely to test efficiency. It is interesting to investigate whether a similar relationship between coverage and performance might exist in the context of spreadsheet debugging. Figure 7 shows a scatterplot for errors corrected versus coverage (blank cells and label cells excluded) including a linear regression model for professionals and students, where the minimum time specified for a cell to be considered inspected/checked was >0.3 seconds. The statistical outliers are circled; these are participants that had standard residuals of <-2 or >2. The $R^2$ value of 0.6199 indicates a moderate correlation.

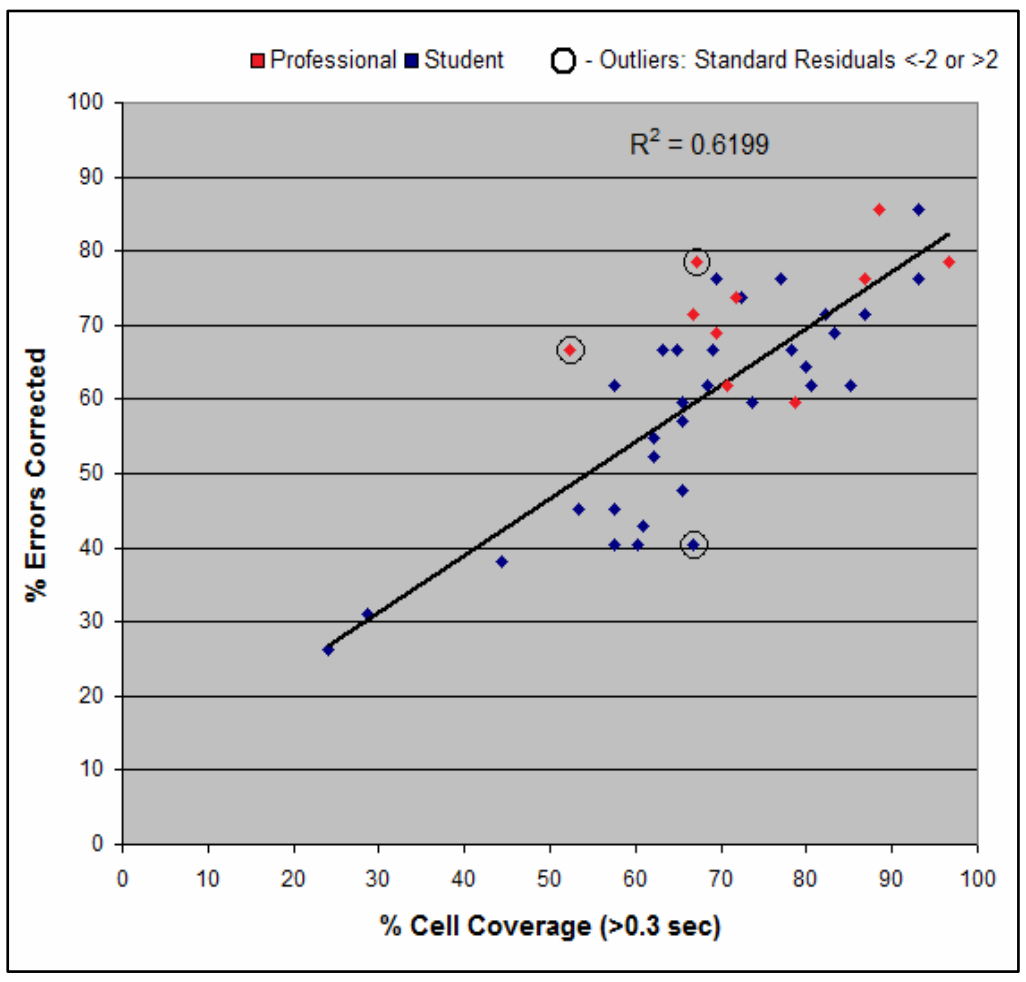

**Figure 7: Errors Corrected/Cell Coverage (outliers circled)**

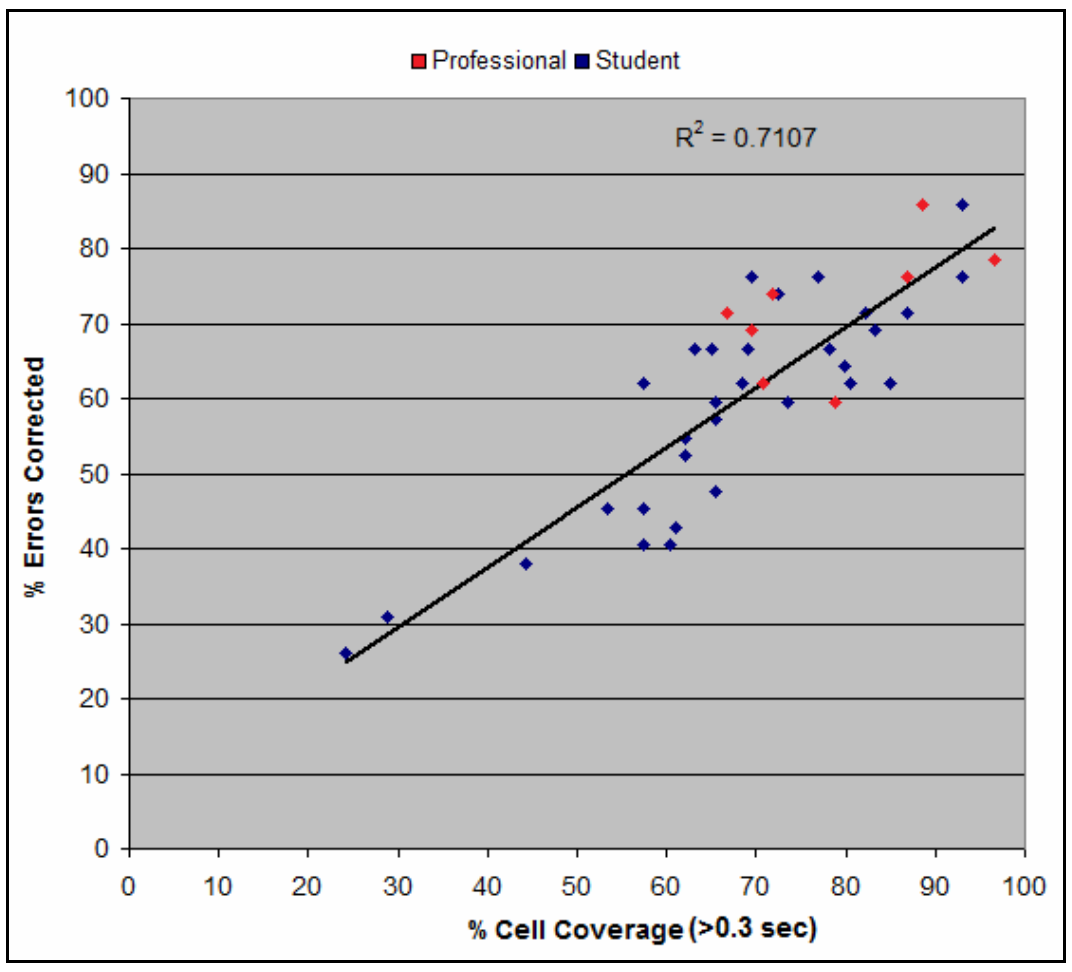

**Figure 8: Errors Corrected/Cell Coverage (outliers removed)**

In an attempt to determine what effect the *statistical outliers* were having on the correlation, the scatterplot in Figure 8 was created. This scatterplot is similar to Figure 7, with the same specified minimum time of 0.3, but with the three *statistical outliers* circled in Figure 7 removed. The $R^2$ value increased to 0.7107.

Although not considered as statistical outliers, the two lowest values on the scatterplot in Figure 7, which are two values closest to the bottom-left of that chart, were considered likely to be having a significant effect on the relationship present. To determine what this effect might be, these two values were removed. This resulting scatterplot (not shown) represented all the available participants, but with the two lowest performers removed. This resulted in an $R^2$ value of 0.4989, which is considered to be a moderate relationship, indicating that the two lowest performers' results were increasing the $R^2$ value by 0.121. Note that the equation of the best fitting regression line for this reduced data set was $y = 0.7701x + 8.1019$, compared with $y = 0.7694x + 8.1536$ for the line in Figure 7, showing that the slope and intercept change little with the removal of the outliers.

## 4. DEBUGGING TOOL DEVELOPMENT

From analysis carried out on the spreadsheet debugging experiment data, see previous section, a correlation was shown to exist between the number of cells inspected and edited, and debugging performance. In traditional software development, the average lines of code inspected and average inspection rates are key metrics used in the code inspection phase of software V&V [Barnard & Price, 1994]. The number of cells inspected during debugging in the spreadsheet paradigm is somewhat equivalent to the lines of code inspected metric used in traditional software V&V. Based on the findings from the spreadsheet debugging experiment, and the use of lines of code inspected metrics in traditional software V&V, a spreadsheet debugging tool was developed that allows the spreadsheet debugger to view those cells that have and have not been inspected. The debugging tool runs concurrently with the T-CAT tool, and uses data gathered by T-CAT to give feedback to users on cells that have been inspected during the debugging process.

### 4.1 Debugging Tool: Cell Coverage Feedback

The debugging tool can be installed as an Excel add-in. Once installed, a button titled ‘Highlight’ is added to each worksheet; see Figure 9. When a user clicks on this button, any cells that have not yet been edited, or selected for a minimum specified time of 0.3 seconds, become highlighted. The user can then inspect the highlighted cells. Again, the minimum specified time of 0.3 seconds was set based on analysis of the minimum time it took participants to inspect a cell, although future versions of the tool might possibly allow for altering of specified time depending on cell inspection speeds. An example of a worksheet in which the Highlight button has been clicked, with any cells that had not been inspected highlighted in grey, can be seen in Figure 9.

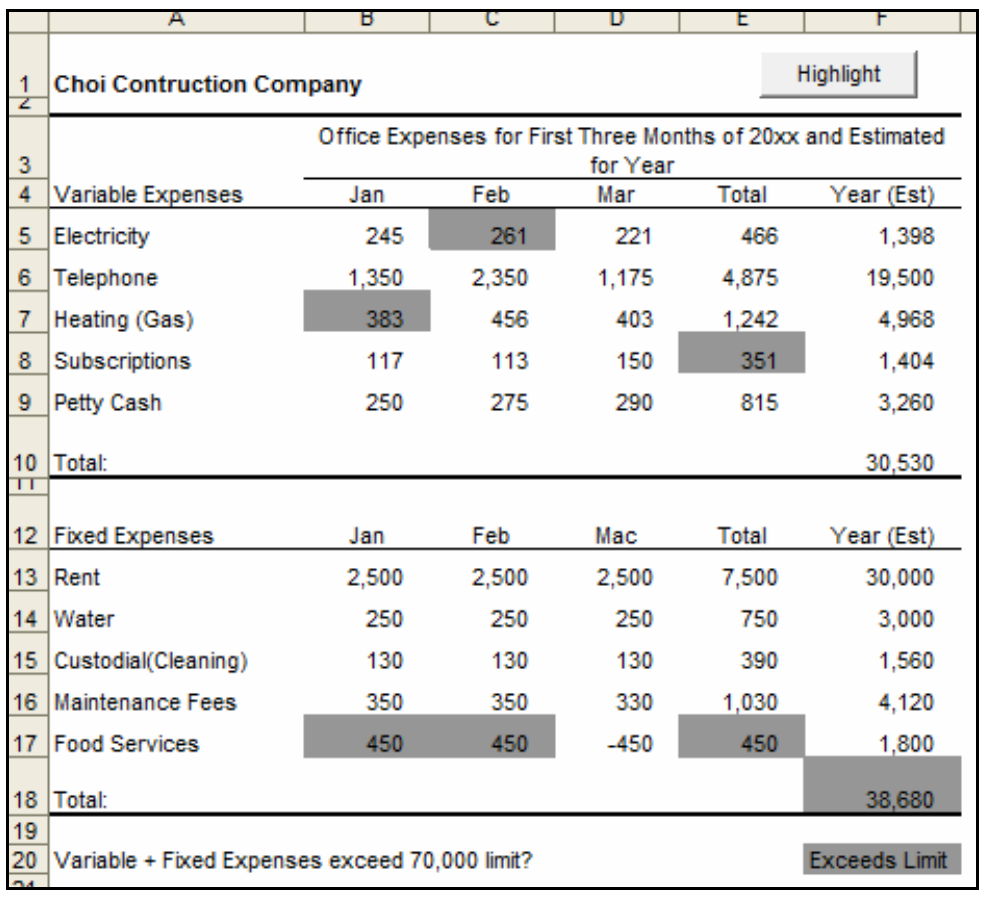

| | A | B | C | D | E | F |
|---|---|---|---|---|---|---|
| 1 | Choi Contruction Company | | | | | Highlight |
| 2 | | | | | | |
| 3 | | Office Expenses for First Three Months of 20xx and Estimated for Year | | | | |
| 4 | Variable Expenses | Jan | Feb | Mar | Total | Year (Est) |
| 5 | Electricity | 245 | 261 | 221 | 466 | 1,398 |
| 6 | Telephone | 1,350 | 2,350 | 1,175 | 4,875 | 19,500 |
| 7 | Heating (Gas) | 383 | 456 | 403 | 1,242 | 4,968 |
| 8 | Subscriptions | 117 | 113 | 150 | 351 | 1,404 |
| 9 | Petty Cash | 250 | 275 | 290 | 815 | 3,260 |
| 10 | Total: | | | | | 30,530 |
| 11 | | | | | | |
| 12 | Fixed Expenses | Jan | Feb | Mac | Total | Year (Est) |
| 13 | Rent | 2,500 | 2,500 | 2,500 | 7,500 | 30,000 |
| 14 | Water | 250 | 250 | 250 | 750 | 3,000 |
| 15 | Custodial(Cleaning) | 130 | 130 | 130 | 390 | 1,560 |
| 16 | Maintenance Fees | 350 | 350 | 330 | 1,030 | 4,120 |
| 17 | Food Services | 450 | 450 | -450 | 450 | 1,800 |
| 18 | Total: | | | | | 38,680 |
| 19 | | | | | | |
| 20 | Variable + Fixed Expenses exceed 70,000 limit? | | | | | Exceeds Limit |

**Figure 9: Debugging Tool Highlighting**

The debugging tool was developed using VBA. When the Highlight button is clicked, the debugging tool is activated. Any current highlighting on the worksheet, that was a result of the Highlight button having previously been clicked, is cleared. The current worksheet name is recorded. Any time-stamped cell activity data that has been recorded by the concurrently running T-CAT tool is retrieved and stored to an array. The array is traversed, and each element of the array (the elements of the array represent a cell selection or cell edit activity) is checked. If the cell (element) in question is a formula cell, or a data cell, and if the cell has not been edited, or has not been selected for >0.3 seconds, then that cell is highlighted. Only those cells from the worksheet on which the Highlight button was clicked can be affected. This process is repeated for all recorded cell selection and cell edit actions. If the Highlight button is clicked again, the highlighting is updated.

### 4.2 Debugging Tool Evaluation Experiment

In order to evaluate the debugging tool, a second experiment was conducted. The sample for this second experiment was made up of 16 fourth year Software Development students. The students were randomly selected and assigned to one of two groups. The 8 participants in the control group, Group-A, debugged the spreadsheet without the tool. The debugging tool was made available to the 8 participants in the test group, Group-B, and instructions were given to them on how to use the tool. The participant worked on similar computers using Microsoft Excel 2000. When the spreadsheets were closed, all of the data relating to any highlighting actions (times Highlight button was clicked, cells affected etc.) was printed to another worksheet. This data could then be analysed to identify how subjects used and interacted with the debugging tool.

### Performance

The overall debugging performance categorised by error type for the eight participants not using the debugging tool, Group-A, and the eight Group-B participants to whom the debugging tool was available can be seen in Figure 10. The test group participants who used the *cell coverage tool* corrected slightly more errors overall, 62%, than the control group, Group-A, correcting 59% of the errors.

| Error Category | Number of Seeded Errors | Group-A (no coverage tool) | Group-B (using coverage tool) |
|---|---|---|---|
| Clerical/Non Material | 4 | 13% | 22% |
| Rules Violation | 4 | 66% | 72% |
| Data Entry | 8 | 66% | 75% |
| Formula | 26 | 63% | 63% |
| **Overall Average** | | 59% | 62% |

**Figure 10: Debugging Performance: Group-A and Group-B (using coverage tool)**

The test group corrected 9% more *clerical* errors, 6% more *rule violation* errors and 9% more *data entry* errors. The same number of *formula* errors were corrected by both groups. However, based on a 5% significance level, there is no statistically significant evidence that the tool aids students to find Clerical/Non-Material, Rules Violation, Data Entry or Formula errors.

Following a detailed examination of the behaviour of Group B participants it was found that 25 more errors were detected and corrected in cells identified by the tool after participants had chosen to use the tool on a sheet. This amounted to a 7.4% increase in error correction overall. Of course, these errors may have been found without the use of the tool. Of the 25 extra errors corrected, 22 were formula errors, 2 were data entry errors and 1 was a clerical error. It is important to note that this analysis must be considered in light of the change in debugging behaviour of the user due to the tool. Some users may consult the tool very regularly and thus most cells will be highlighted and any errors found may be considered as ones identified by the tool. This work next investigates the impact of the tool on cell coverage and the behaviour of the debugger.

**Impact of the Tool on Cell Coverage**

The cell coverage for participants in Group A and B can be seen in the boxplot in Figure 11. The test group using the cell coverage tool achieved a significantly higher cell coverage rate than the control group. The minimum time specified for a cell to be considered inspected was >0.3 seconds. The overall average cell coverage for Group B was 90%, 26% higher than Group A, who achieved an overall average cell coverage of 64%.

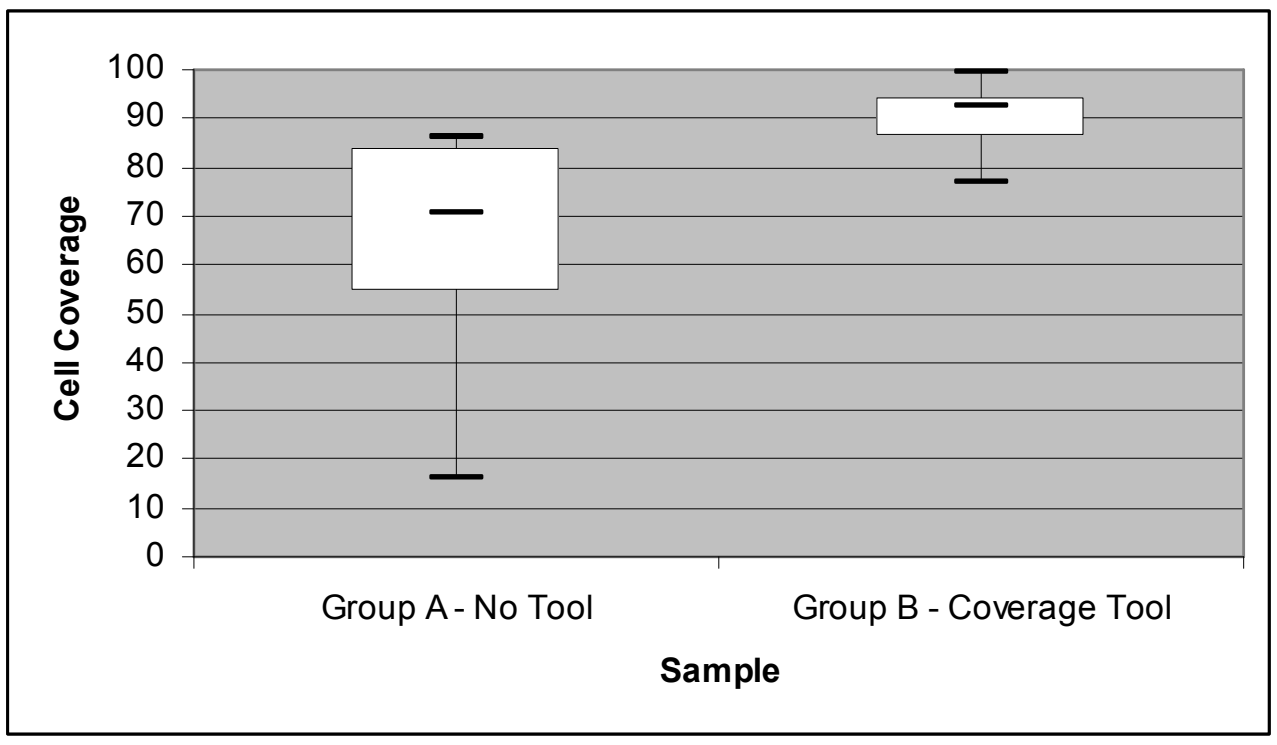

**Figure 11: Cell Coverage Boxplot: Group A and B.**

**Subject Behaviour While Using the Debugging Tool**

Although the 8 Fourth year Software Development students who took part in the experiment were given the same instructions for using the *cell coverage debugging tool* during the debugging process, there was some differences in the way each student used the tool. The following figures represent two of the eight examples of participant behaviour in using the tool and serve to investigate the impact of the tool on debugging behaviour.

Figure 12 shows the behaviour of one of the participants using the debugging tool. What can be seen is that the participant clicked the 'Highlight' button on the 'Payroll' sheet after 3285 seconds and 8 cells became highlighted. This process was repeated for the 'Office Expenses' and 'Projections' sheet. The participant in this case had inspected almost all of the cells in the spreadsheet, and when the highlight buttons were clicked, only a few cells became highlighted. In this case, the participants debugging behaviour does not seem to have been affected by the tool in that they waited until they had completed the entire task before utilising the tool, and the few cells that were highlighted could be checked by the participant as desired.

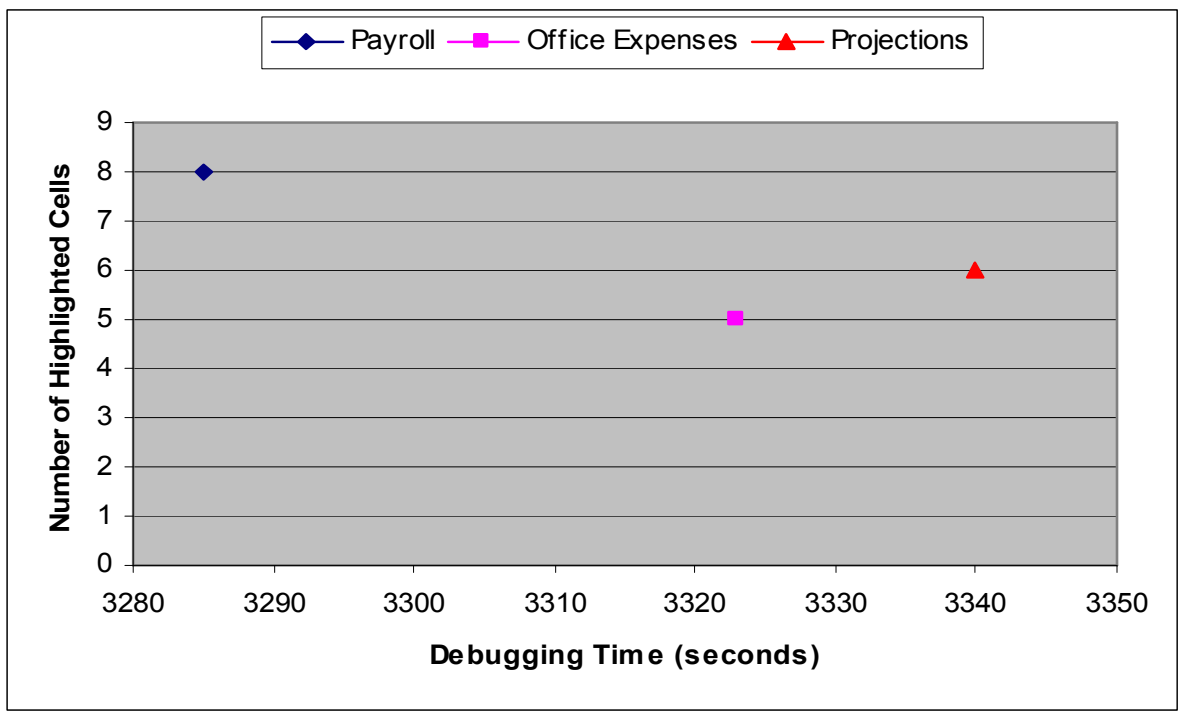

**Figure 12: Debugging Tool Behaviour: Subject 1**

The behaviour of another participant, as represented in Figure 13, appears to involve the use of the cell coverage debugging tool to a greater extent. The figure shows that the highlight button on the 'Payroll' sheet was clicked 4 times. It also shows that the participant re-inspected the 'Payroll' and 'Office Expenses' sheets, using the coverage tool each time. Using the tool the participant rechecked the first two sheets before finalising the task.

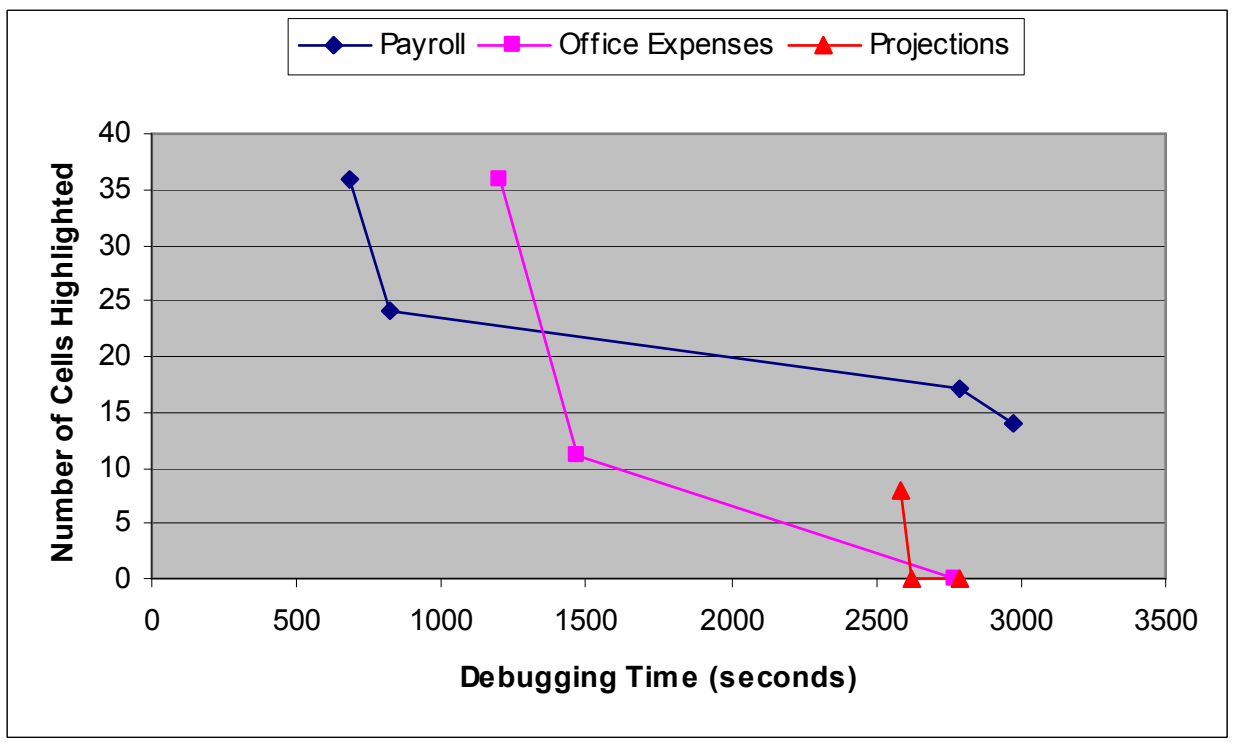

**Figure 13: Debugging Tool User Behaviour: Subject 3**

Overall, it was found that the subjects interacted with the debugging tool in different ways and to different extents. Hence, it may be beneficial to explore and develop a supporting process which would guide the debugger in the use of the tool in the most advantageous manner. Research is ongoing in this regard.

## 5. CONCLUSION

This paper details the design and successful implementation of a custom built data acquisition method and tool (T-CAT) for the spreadsheet research area. The T-CAT tool was an integral part of a spreadsheet debugging experiment undertaken by the authors. Based on findings from analysis of debugging behaviour recorded by T-CAT, a debugging tool was developed by the authors, and its effects on debugging performance were investigated by means of a controlled experiment. Although the debugging tool was not as effective as was initially hoped, the debugging performance of subjects using the tool was slightly higher overall, and overall cell coverage was significantly higher. The T-CAT data acquisition tool was used in both experiments, and along with increasing the authors understanding of end-users' natural debugging behaviour, it also aided in the design and evaluation of the debugging tool.